# Cross-correlation image analysis for real-time single particle tracking

L. R. Werneck*,[1] C. Jessup,[2] A. Brandenberger,[2] T. Knowles,[3]
C. W. Lewandowski,[2,4] M. Nolan,[2] K. Sible,[5,6,7] Z. B. Etienne,[1,5,6] and B. D'Urso[2]

[1]*Department of Physics, University of Idaho, Moscow, ID 83843, USA*
[2]*Department of Physics, Montana State University, Bozeman, MT 59717, USA*
[3]*Department of Mathematics, West Virginia University, Morgantown, WV 26506, USA*
[4]*Space Dynamics Laboratory, Albuquerque, NM 87106, USA*
[5]*Department of Physics and Astronomy, West Virginia University, Morgantown, WV 26506, USA*
[6]*Center for Gravitational Waves and Cosmology, Chestnut Ridge Research Building, Morgantown, WV 26506, USA*
[7]*Department of Computer Science and Engineering, University of Notre Dame, South Bend, IN 46556, USA*
(*The author to whom correspondence may be addressed: wernecklr@gmail.com)
(Dated: June 3, 2024)

Accurately measuring the translations of objects between images is essential in many fields, including biology, medicine, chemistry, and physics. One important application is tracking one or more particles by measuring their apparent displacements in a series of images. Popular methods, such as the center-of-mass, often require idealized scenarios to reach the shot noise limit of particle tracking and are, therefore, not generally applicable to multiple image types. More general methods, like maximum likelihood estimation, reliably approach the shot noise limit, but are too computationally intense for use in real-time applications. These limitations are significant, as real-time, shot-noise-limited particle tracking is of paramount importance for feedback control systems. To fill this gap, we introduce a new cross-correlation-based algorithm that approaches shot-noise-limited displacement detection and a GPU-based implementation for real-time image analysis of a single particle.

## I. INTRODUCTION

Tracking the motion of particles in a sequence of images over time is a common practice in multiple research fields, from biology [1–5] to chemistry [6–8], medicine [9, 10], and physics [11–18]. The methods employed can be divided into those suitable for real-time [9, 11–13] and offline (not real-time) use [1–8, 14–19]. They can be further divided into those that measure the particle displacement with shot-noise-limited precision [1] and those that do not [2–9, 11–19].

Existing offline tracking techniques for single particles include the nonlinear least-squares fit for shot-noise-limited localization using a particle's point spread function [1] and Lorenz-Mie scattering theory for sub-pixel resolution [19]. Other sub-pixel precision methods include nonlinear least squares for fitting single fluorophores' intensity distributions to two-dimensional Gaussian profiles [3, 4], employing the center-of-mass method (also known as the moment method) to track single particles in dusty plasma [15–18], and a non-iterative fit for determining the optimal center of radial symmetry of an imaged particle [2].

Alternatively, the azimuthal symmetry of a particle's image and the shift property of the Fourier transform can be exploited to track the position of the particle [20, 21]. In addition, a convolutional neural network is employed for noise tolerance in environments with varying signal-to-noise ratios, consistently tracking particle positions across a spectrum from low (∼1) to high (up to ∼40) signal-to-noise ratios [7].

Fewer methods are available for real-time analysis, and these are typically application dependent. In medicine and biotechnology, for example, a technique leveraging artificial intelligence performs real-time tracking and feedback control of multiple particles to determine their size and location for *in vitro* diagnostics [9]. In physics, applications often employ the center-of-mass algorithm to perform real-time feedback control of a single particle based on its position. This is a computationally inexpensive technique that performs well when the input images consist of a bright (or dark) spot against a relatively featureless dark (or bright) background [11, 12]. While the center-of-mass algorithm can reach the shot noise limit of particle tracking under ideal conditions, this method is susceptible to biases when background noise or light is present [5, 13, 14].

Besides the center-of-mass algorithm, the cross-correlation (CC) method is a popular choice for offline image registration [12, 22–26]. This method measures the apparent displacement between two images via the position of the maxima of the CC between them, and can be modified to yield sub-pixel displacements [27] at the cost of increased computational complexity. Several open-source implementations of the CC method exist, but these are typically not designed with real-time analysis in mind and generally demand enormous computational resources to achieve the necessary processing speed. Further, the CC method does not generally target shot-noise-limited accuracy, which is crucial for feedback systems where quantum-limited control is desirable [28, 29].

While many particle tracking applications use camera-based detection, quadrant photodetectors, discrete photodiodes, and balanced detectors are commonly used in levitated optomechanics [30–37] and approach shot-noise limited detection in real time, but only with small displacements [28]. A hybrid detection scheme using photodiodes for rotation detection with subsequent analysis of high-speed CMOS camera images to measure translational motion has been used to study the coupled rotation and translation of levitated birefringent particles [35–37]. Alternatively, event-based imaging has been demonstrated for tracking silica microspheres in a Paul trap over displacements exceeding $100\,\mu\mathrm{m}$ with $30\,\mathrm{nm}/\sqrt{\mathrm{Hz}}$ sensitivity at 1 kHz acquisition rates [38]. This technique uses pixel arrays with contrast detectors that trigger upon intensity changes beyond a preset threshold. The changed pixels feed into a tracking algorithm that analyzes each frame to detect object motion. Acquisition rates could surpass 1 GHz [39], potentially enabling real-time feedback control [38]. However, event-based imaging requires specialized hardware and further development of tracking algorithms [38, 39].

In this paper we address the gap in established techniques by introducing real-time, CC-based image analysis methods that approach the shot noise limit of accuracy for single particle tracking. We demonstrate a real-time adaptation of the uniformly-weighted CC (CC-U) method of [12], as well as the new real-time shot-noise-weighted CC (CC-SN) method. We also provide an open-source implementation of these algorithms, demonstrating their real-time analysis capabilities by tracking a microsphere levitated in a magneto-gravitational trap [29], which is our primary application of interest. Our implementation is suitable for commercial off-the-shelf graphics processing units (GPUs) and can analyze hundreds of images per second.

The remainder of this paper is organized as follows. Section II introduces our novel cross-correlation-based image analysis method. In Sec. III, we detail the numerical implementation of this method and discuss its variations for both real-time and offline analyses. Section IV presents our results. First, we apply the method to various types of synthetic particle data, comparing its performance against other widely used methods in the literature and showing that it reliably approaches the shot noise limit. Then we use our method on experimental data, tracking a magnetically levitated particle using both the real-time and offline versions of the algorithm. Finally, in Sec. V, we conclude and discuss future research directions.

## II. BASIC APPROACH

A common method to find the location $\vec{R}$ of a bright spot on a dark background is the center-of-mass (CM) calculation, $\vec{R} = \sum_{\vec{r}} \vec{r} I(\vec{r}) / \sum_{\vec{r}} I(\vec{r})$, with $\vec{r}$ denoting the location of a pixel in an image $I(\vec{r})$. In the ideal case of a nearly featureless background, this method can approach the shot noise limit [28]. However, it is highly affected by background light and image boundaries [13]. A more statistically robust approach, maximum likelihood estimation (MLE), adjusts parameters in a fit distribution to maximize the likelihood that the data comes from the distribution, assuming Poisson statistics. If Gaussian statistics are used as an approximation, we can instead minimize $\chi^2$ relative to the displacement $\vec{r}_0$ between two images, as expressed in the following equation:

$$\chi^2 = \sum_{\vec{r}} \left[ \frac{I(\vec{r} - \vec{r}_0) - E(\vec{r})}{\sigma(\vec{r}, \vec{r}_0)} \right]^2, \quad (1)$$

where $I(\vec{r} - \vec{r}_0)$ is the image $I(\vec{r})$ translated by $\vec{r}_0$ (assuming periodic boundary conditions), $E(\vec{r})$ is the reference image (RI; see Sec. III), and $\sigma(\vec{r}, \vec{r}_0)$ models the noise in the images. Minimizing $\chi^2$ directly is computationally intensive, so more efficient strategies, such as cross-correlation (CC), are generally preferred.

The uniformly-weighted CC (CC-U) method, for example, assumes that $\sigma(\vec{r}, \vec{r}_0)$ has a constant value $\sigma$. Under this assumption, Eq. (1) simplifies to:

$$\chi^2 = \frac{1}{\sigma^2} \sum_{\vec{r}} \left[ I(\vec{r} - \vec{r}_0)^2 - 2I(\vec{r} - \vec{r}_0)E(\vec{r}) + E(\vec{r})^2 \right]. \quad (2)$$

Since the sums over $I(\vec{r} - \vec{r}_0)^2$ and $E(\vec{r})^2$ span the entire images, they are independent of $\vec{r}_0$. We may then write

$$\chi^2 = C - \frac{2}{\sigma^2} \sum_{\vec{r}} I(\vec{r} - \vec{r}_0)E(\vec{r}), \quad (3)$$

where $C$ is some constant independent of $\vec{r}_0$. The displacement $\vec{r}_0$ is then obtained by finding the value of $\vec{r}_0$ that minimizes the right-hand side of Eq. (3), which coincides with the position of the maximum of the CC between the shifted image $I(\vec{r} - \vec{r}_0)$ and the RI $E(\vec{r})$, i.e.,

$$\sum_{\vec{r}} I(\vec{r} - \vec{r}_0) E(\vec{r}) = \mathscr{F}^{-1}\left[ \mathscr{F}[I] \otimes \overline{\mathscr{F}[E]} \right], \quad (4)$$

where $\mathscr{F}$ is the Fourier transform, $\mathscr{F}^{-1}$ its inverse, $\otimes$ denotes element-wise multiplication, and the overline denotes complex conjugation.

For CMOS cameras with adequate light levels, photon shot noise is often the dominant noise source. The noise in each pixel then follows a Poisson distribution, which can be approximated by a Gaussian distribution with a standard deviation equal to the square root of the photon count in the pixel. In this case, a shot-noise-weighted version of the CC method (CC-SN) is more appropriate. This method assumes the RI provides an estimate for the *average* pixel values of the images in a dataset, i.e.,

$$\sigma(\vec{r}, \vec{r}_0) = \sqrt{E(\vec{r})}. \quad (5)$$

For nearly zero pixel values, the square root poorly approximates the uncertainty of the Poisson distribution. Also, negative pixel values may arise when subtracting the average background from images. To compensate, we add a constant offset $\varepsilon$ to the new images (see Sec. III for how this affects the RI), with larger offset values de-emphasizing darker pixels.

Adding the offset to the new image and using Eq. (5) in Eq. (1), we find

$$\chi^2 = \sum_{\vec{r}} \left\{ E(\vec{r}) - 2I'(\vec{r} - \vec{r}_0) + \left[ I'(\vec{r} - \vec{r}_0) \right]^2 / E(\vec{r}) \right\}, \quad (6)$$

where $I' \equiv I + \varepsilon$. As with Eq. (2), the first two sums in Eq. (6) are independent of $\vec{r}_0$, and the minimization of the right-hand side of Eq. (6) only depends on the CC between $I'^2$ and the reciprocal of $E$. Therefore, the displacement $\vec{r}_0$ that minimizes Eq. (6) is determined by the position of the maximum of

$$\sum_{\vec{r}} \left[ I'(\vec{r} - \vec{r}_0) \right]^2 / E(\vec{r}) = \mathscr{F}^{-1}\left[ \mathscr{F}[I'^2] \otimes \overline{\mathscr{F}[1/E]} \right]. \quad (7)$$

## III. NUMERICAL METHODS

The position of the maximum of the CC provides an integer pixel estimate of the displacement between two images. Following [22, 27], we define a 1.5-pixel square region centered on the maximum of the CC and sample it with $1.5u$ points in both the horizontal and vertical directions, where $u$ is the upsampling factor. When $u = 256$, for example, the 1.5-pixel square region is sampled with resolution $384 \times 384$. Looking at just this small region produces massive efficiency gains, as the effective resolution around the square is the same as if the original image had $256^2 \times 256^2$ pixels. Data from the element-wise product on the right-hand sides of Eq. (4) or (7) (depending on the method) is then interpolated into the upsampled region, followed by an inverse discrete Fourier transform to generate an upsampled CC. The position of the maximum of the upsampled CC yields a sub-pixel estimate of the displacement.

The CC-U and CC-SN methods can be used to analyze images both in real-time and with an iterative method that uses past and future images, with different approaches for updating the RI in each case. The following discussion focuses on the CC-U method, but it can be easily adapted for the CC-SN method by replacing $I_i \to I_i'^2$ and $E_n \to 1/E_n$.

### A. Offline Analysis

In the offline case [12], we iteratively update the RI as the average in Fourier space of all images within a dataset $\{I_0, I_1, \ldots, I_{N-1}\}$. For each complete pass through the images, the Fourier transform of the RI at the $n$-th iteration, $\mathscr{F}[E_n]$, is computed using

$$\mathscr{F}[E_n] = \frac{1}{N} \sum_{i=0}^{N-1} \mathscr{F}\left[I_i(\vec{r} - \vec{r}_0)\right], \quad (8)$$

where $\mathscr{F}[E_0] = \mathscr{F}[I_0]$. For a given iteration, computing the displacement of each image in the dataset with respect to the constant RI is an embarrassingly parallel operation. Thus, the dataset can be split across multiple processing units for significant speed-up.

### B. Real-time Analysis

Real-time analysis is inherently a sequential process that requires continuously updating the RI as each new image is processed. The computational expense is large, as both the displacement and the RI must be updated at least at the camera's frame rate. As each new image is processed, the RI is updated using a low-pass filter,

$$\mathscr{F}[E_i] = a_0 \mathscr{F}\left[I_i(\vec{r} - \vec{r}_0)\right] + b_1 \mathscr{F}[E_{i-1}], \quad (9)$$

where $\mathscr{F}[E_0] = \mathscr{F}[I_0]$, $a_0 = 1 - e^{-T/\tau}$, $b_1 = e^{-T/\tau}$, $T$ is the inverse of the camera's frame rate, and $\tau$ is the filter time constant (typically, $\tau \gg T$).

This method may experience drift over time as a result of the gradual accumulation of numerical errors, which can displace the particle in the RI from its initial position. Similarly, drift in the particle's actual position can occur over time due to changes in its equilibrium position. Although these drifts usually have no significant consequences, they can be inconvenient, and it is often desirable to eliminate them from the detected motion, retaining only the faster particle oscillations.

A natural solution is to subtract the (possibly drifting) equilibrium position, as determined by a low-pass filter. Given the $n$-th raw displacement $\vec{r}_n$, we calculate the filtered displacement value $\vec{r}_n' = \vec{r}_n - \vec{R}_n$, where $\vec{R}_n$ is the subtracted offset calculated by passing the raw data through a low-pass filter,

$$\vec{R}_n = a_0 \vec{r}_n + b_1 \vec{R}_{n-1}, \quad (10)$$

and $\vec{R}_0 = a_0 \vec{r}_0$.

If there is a concern that Eq. (10) may be too aggressive in correcting drift and could inadvertently remove an important signal, the shift in the offset can be limited to a fraction $\alpha$ of the resolution of the CC analysis (the inverse of the upsampling factor $u^{-1}$). In this case, the drift correction effectively biases the round-off error in the CC toward a zero average displacement, and the change in the correction offset can be written as

$$d\vec{R}_n = \left(a_0 \vec{r}_n + b_1 \vec{R}_{n-1}\right) - \vec{R}_{n-1} = a_0 \vec{r}_n + (b_1 - 1)\vec{R}_{n-1}, \quad (11)$$

and

$$\vec{R}_n = \begin{cases} \vec{R}_{n-1} + (\alpha/u) d\hat{R}_n & \text{if } |d\vec{R}_n| > \alpha/u, \\ \vec{R}_{n-1} + d\vec{R}_n & \text{otherwise}, \end{cases} \quad (12)$$

where $d\hat{R}_n = d\vec{R}_n / |d\vec{R}_n|$. A typical choice of parameter is $\alpha = 0.5$, ensuring that drift corrections are smaller than the expected displacement resolution.

## IV. RESULTS

We perform tests with synthetic data, where each synthetic image comprises a Gaussian function integrated over each pixel, representing either a bright or dark particle on a dark or bright background, with Poisson shot noise added (see Fig. 1). To ensure that these synthetic images are experimentally relevant, we match their maximum pixel intensity to the saturation capacity of the CMOS sensor used in our experiments (10,700 electrons for a Sony IMX250). When adding dark noise to the images, we also match it to the dark noise of the sensor (2.4 electrons). For each method and choice of parameters, we analyze data sets consisting of 1,000 images of size $512 \times 512$, whose displacements are randomly distributed but exactly known. For the CC-U and CC-SN methods, we fix the upsampling factor to $u = 512$.

For the bright spot on a dark background, we consider both the case without and with dark noise, where a constant value is added to the entire image before the Poisson noise is introduced. This means that subtracting the mean background value from the image with dark noise does not yield the image without dark noise. For the dark spot on the bright background, we only consider the case with added dark noise.

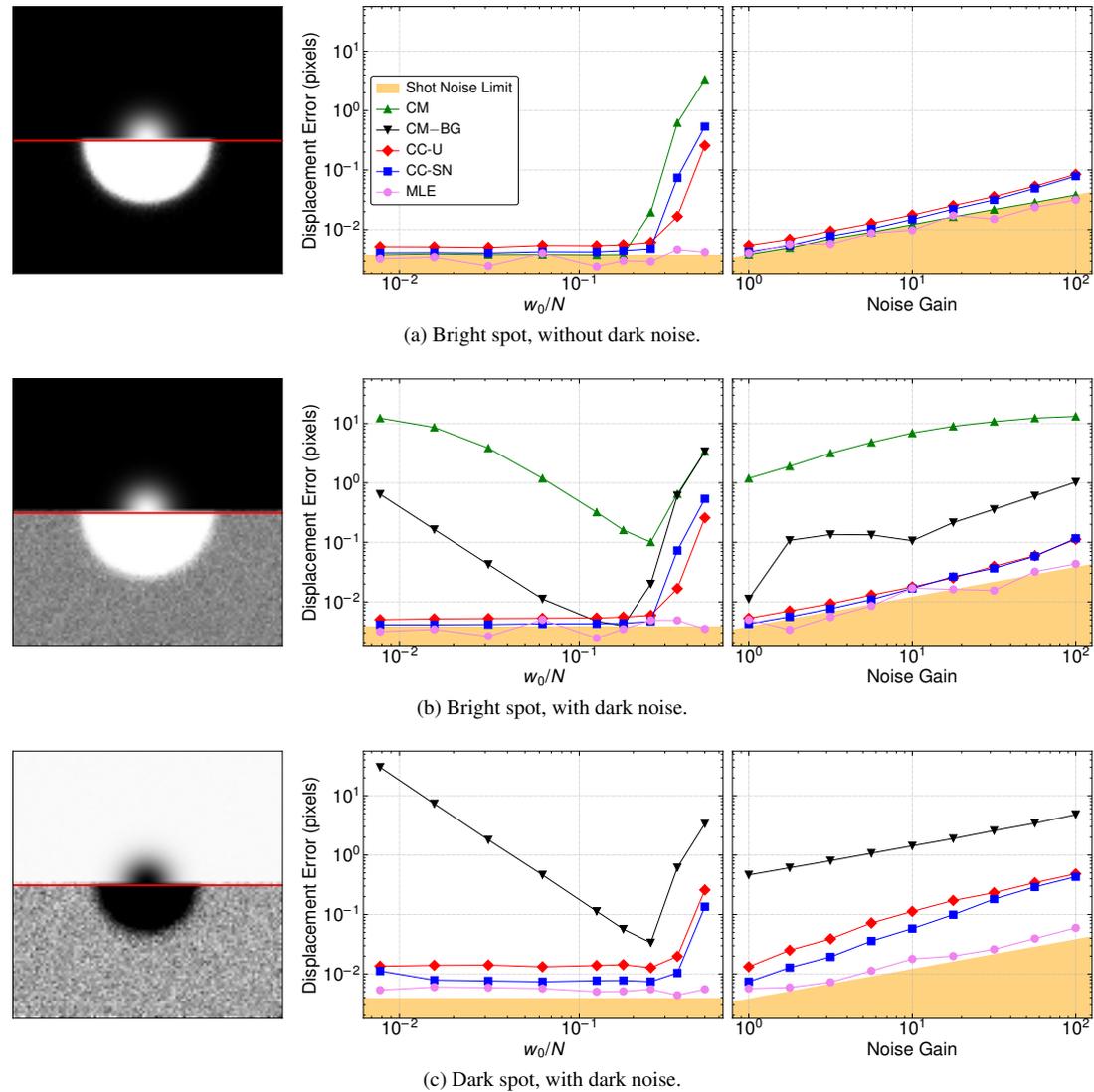

FIG. 1. Gaussian particle tracking study: comparison of various image analysis methods in the presence and absence of dark noise. The legend indicates the particular method used: center-of-mass (CM), CM subtracting the mean background (CM − BG), uniform cross-correlation (CC-U), shot-noise-weighted cross-correlation (CC-SN), and maximum likelihood estimation (MLE). The Gaussian width as a fraction of the image width ($w_0/N$) and the noise gain are used as comparison metrics on the center and right columns, respectively. The left column shows samples of the analyzed images: the top half of each panel presents the original, and the bottom half adjusts the pixel values (through rescaling, offsetting, and saturating) to highlight the presence or absence of background noise.

We compare the uniformly-weighted (CC-U) and shot-noise-weighted (CC-SN) cross-correlation methods described in Sec. II with the standard center-of-mass (CM) and maximum likelihood estimation (MLE) methods, using the real-time approach described in Sec. III. When using the MLE method, we provide the exact displacement as the initial value for the iterative optimization algorithm. By doing so, the MLE method produces results that represent the best achievable outcomes from numerical methods, offering a benchmark against which other methods can be compared.

The performance of each method in recovering displacements for a bright spot in the absence of dark noise is shown in the center column of Fig. 1(a). In this case, all methods exhibit roughly the same performance as MLE, effectively reaching the shot noise limit for the standard deviation of the position error, $w_0/(2\sqrt{N_\gamma})$, where $N_\gamma$ is the total number of photons detected in the Gaussian [28]. This is valid as long as the particle size (proportional to the Gaussian width $w_0$) does not constitute a large fraction of the image. In this idealized context, the CM algorithm is the preferred choice, given its simplicity and low computational cost.

We then perform the same analysis for the more realistic case of a bright spot with dark noise, and the results are displayed in the center column of Fig. 1(b). In this scenario, the CM algorithm is used to analyze both the original images and the images with the exact mean background subtracted (CM − BG). While subtracting the mean background substantially enhances the CM method's performance, it still falls short of reaching the shot noise limit, except for a small range of $w_0$, a feature which is used in [13]. On the other hand, both the CC-U and CC-SN methods yield results comparable to those obtained using the MLE method and are therefore preferred due to their significantly lower computational cost. As in the previous case, the error for the CC-U method is slightly larger than those for the CC-SN and MLE methods. Ignoring the largest two values of $w_0/N$, we find that the MLE method produces results that are, on average, within 31 % of the shot-noise limit, while the same average for the CC-SN and CC-U are 57 % and 97 %, respectively.

Finally, the center column of Fig. 1(c) contains results from analysing images consisting of a dark spot in the presence of dark noise. When using the CM method, we subtract the image from the exact mean background value to approximate a bright spot on a dark background. Nevertheless, the displacement errors for the CM method are still about an order of magnitude above the shot noise limit. While the CC-U method exhibits larger errors than the MLE method, they are still notably smaller than those associated with the CM method. The errors for the CC-SN method remain comparable to those for the MLE method.

To evaluate how well the methods perform under varying signal-to-noise ratios, we have repeated our analysis by fixing $w_0 = 32$—the central point of the curves in the center column of Fig. 1—and varying the noise gain of the images. The results are shown in the right column of Fig. 1. We observe that the MLE method outperforms the others, followed closely by the CC-SN and CC-U methods, even at lower signal-to-noise ratios. We have empirically determined that setting the constant offset $\varepsilon$ (see Sec. II) to 200 times the noise gain yields optimal results when using the CC-SN method. Smaller values of $\varepsilon$ lead to worse displacement errors, while for larger values the CC-SN method's results approaches those of the CC-U method.

The processing speed of different algorithms is critically important for real-time particle tracking. To assess it, we conducted performance benchmarks of our CC-SN algorithm on a high-end personal computer equipped with an AMD Ryzen 9 5950X 16-core CPU and an NVIDIA RTX 3080 GPU. As shown in Table I, we analyzed the processing speed for various image sizes and observed a consistent drop as the upsampling factor $u$ increased. This decline is expected, as the upsampling algorithm is the most computationally intensive component of the code. Unsurprisingly, the fastest implementation is the CUDA one.

The difference in processing speed between the CC-U and CC-SN algorithms is negligible. However, the MLE method is significantly slower, taking, on average, more than 10 s to analyze a 512×512 image. To highlight this contrast, consider the slowest analysis (upscale factor $u = 1024$) for the same image size in our CUDA implementation: 31.1 images per second, which is more than 300 times faster than MLE.

Even our Python implementation—which is unsuitable for real-time analysis since it can only process 8.8 images per second—is about 90 times faster than MLE in the worst-case scenario. These comparisons underscore the superior efficiency of the CUDA implementation over the MLE method.

TABLE I. Benchmark results for different CC-SN algorithm implementations on an AMD Ryzen 9 5950X CPU (Python & C) and NVIDIA RTX 3080 GPU (CUDA).

| Image Size | Upsampling | Implementation | Images/s |
|---|---|---|---|
| 256×128 | 1024 | Python | 15.5 |
| | 1024 | C | 19.5 |
| | 1024 | CUDA | 91.9 |
| | 512 | CUDA | 325 |
| | 256 | CUDA | 811 |
| | 128 | CUDA | 1636 |
| 256×256 | 1024 | Python | 13.3 |
| | 1024 | C | 14.1 |
| | 1024 | CUDA | 67.0 |
| | 512 | CUDA | 189 |
| | 256 | CUDA | 432 |
| | 128 | CUDA | 981 |
| 512×256 | 1024 | Python | 11.7 |
| | 1024 | C | 12.2 |
| | 1024 | CUDA | 58.6 |
| | 512 | CUDA | 168 |
| | 256 | CUDA | 322 |
| | 128 | CUDA | 670 |
| 512×512 | 1024 | Python | 8.8 |
| | 1024 | C | 9.8 |
| | 1024 | CUDA | 31.1 |
| | 512 | CUDA | 93 |
| | 256 | CUDA | 169 |
| | 128 | CUDA | 348 |
| 1024×128 | 1024 | Python | 9.8 |
| | 1024 | C | 13.2 |
| | 1024 | CUDA | 85.3 |
| | 512 | CUDA | 227 |
| | 256 | CUDA | 360 |
| | 128 | CUDA | 717 |

### A. Experimental data

We show the CC-SN method can be used for real-time analysis by tracking a borosilicate glass microsphere levitated in a magneto-gravitational trap in high vacuum. The microsphere is back-illuminated with collimated light from a pulsed 660 nm LED and imaged on a CMOS camera (Basler acA2440-75 μm), giving the particle the appearance of a dark disk on a bright background with a bright spot in the center (see Fig. 2; see also [12, 29]).

Applying a voltage across the vertical gap of the trapping region creates an electric field that exerts additional force to suspend the microsphere, which has a diameter of approximately 68 μm [12]. The recorded data consists of 256 × 128 pixel images of the microsphere, corresponding to a field of view of approximately 310 × 150 μm, with 12 bits per pixel for ten minutes at a rate of 470 Hz, which is near the maximum frame rate supported by the camera's USB3 interface.

The CM algorithm proves unsuitable for tracking the particle due to the complex structure of the microsphere's image, characterized by non-Gaussian features such as diffraction rings along the edges and a prominent bright spot at the center. These features demand a more sophisticated approach for accurate tracking, such as CC-SN. Although CC-SN is significantly more computationally demanding than CM, our implementation makes it GPU-compatible. This enables us to measure the displacement of the particle at high resolutions and framerates in real-time using a GeForce RTX 3080 Ti GPU, while simultaneously recording the images. For comparison, we perform an offline analysis using the CC-SN algorithm and our iterative method described in Sec. III up to the fifth RI. In both cases, the upsampling factor is set to 256.

We find excellent agreement between the real-time CC-SN method and the previously demonstrated offline CC-U method, with differences of less than 0.15 pixels for the entire data set, as shown in Fig. 3. We note that, for the purposes of this comparison, we disable the slow drift correction algorithm described in Sec. III for both methods. The small drift in the residual of the displacements calculated by these methods, observed in the bottom panels of Fig. 3, is expected and attributed to the fact that the real-time algorithm drifts numerically.

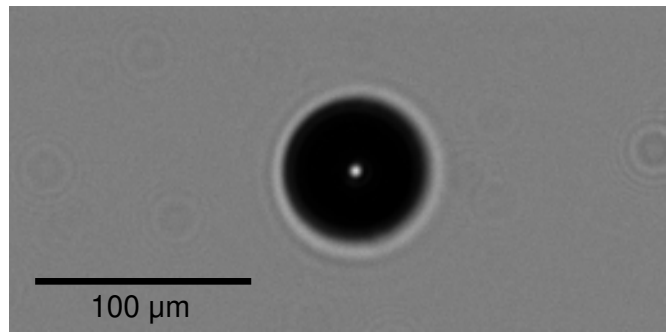

FIG. 2. Experimental data: backlit, levitated borosilicate glass microsphere. Artifacts in the image background are from dust on the optics and window of the vacuum chamber.

## V. CONCLUSIONS

In this paper, we introduced a shot-noise-weighted cross-correlation (CC-SN) method for determining image displacements near the shot noise limit. The method was tested on several different images, including a bright spot on a dark background, both with and without dark noise, a dark spot on a bright background with shot noise, and on complex experimental data. We found that the CC-SN method is superior to the uniformly weighted cross-correlation (CC-U) method, with errors comparable to those obtained using the maximum likelihood (MLE) method, while being significantly more computationally efficient than MLE.

Since CC-SN can be used for real-time image analysis, it can be implemented in feedback control systems [29, 40, 41]. To this end, we have implemented the CC-SN method to track the position of a microsphere levitated in a magneto-gravitational trap in high vacuum, in real-time. Looking ahead, we plan to apply the method for feedback cooling of the motion of a particle with shot-noise-limited precision, as well as for state preparation and tracking for a new measurement of the gravitational constant $G$ with a levitated particle [41].

Using this method for real-time analysis requires an implementation of the algorithms discussed in this paper that is capable of processing images at a rate at least equal to the camera's frame rate. To this end, we provide an open-source toolkit for real-time image analysis called RETINAS [42]. The toolkit contains the methods described in Sec. II implemented in C and Python—as well as a GPU-capable implementation written in CUDA—all of which can be accessed via a user-friendly Python interface. Reference [42] contains both the code and documentation on how to use the code and reproduce all the results shown in this paper.

The real-time CC-SN method described in this paper represents a significant advancement in real-time image analysis, bridging a critical gap in current particle tracking algorithms. With shot-noise-limited accuracy, the technique is particularly well-suited to levitated optomechanics, where quantum-limited detection and feedback based on particle motion are targeted.

### ACKNOWLEDGMENTS

This material is based upon work supported by the National Science Foundation under grant numbers 1806596, 1912083, 1950282, 2011783, and 2227079. BD gratefully acknowledges a Block Gift from the Coherent / II-VI Foundation. This research made use of the resources of the High Performance Computing Center at Idaho National Laboratory, which is supported by the Office of Nuclear Energy of the U.S. Department of Energy and the Nuclear Science User Facilities under contract number DE-AC07-05ID14517. ChatGPT was used on occasion for language and readability improvements. The authors assume full responsibility for the contents of this work. Data presented in this paper are available in Ref. [43].

[1] Y.-H. Lin, W.-L. Chang, and C.-L. Hsieh, Opt. Express **22**, 9159 (2014).
[2] R. Parthasarathy, Nature Methods **9**, 724 (2012).
[3] C. M. Anderson, G. N. Georgiou, I. E. G. Morrison, G. V. W. Stevenson, and R. J. Cherry, Journal of Cell Science **101**, 415 (1992), https://journals.biologists.com/jcs/article-

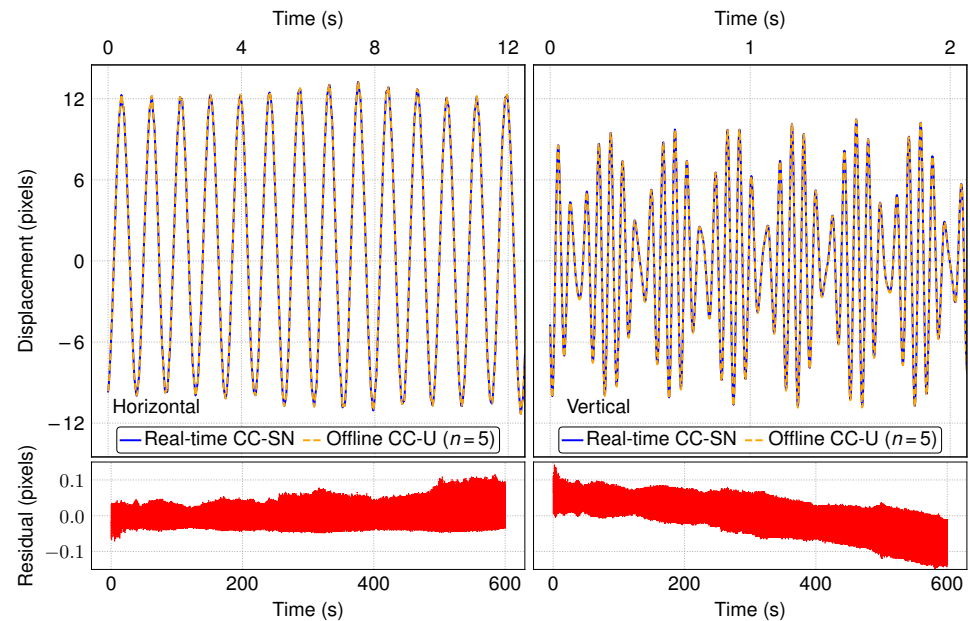

FIG. 3. Comparison between real-time CC-SN and the previously demonstrated offline CC-U algorithms at the fifth reference image for experimental data. The calibration is 1.2 μm/pixel. **Top**: Displacements $r_0$. The apparent beating in the detected vertical motion is due to coupling with the transverse ($x$) motion (see [12, 29]) separated in frequency by ∼3 Hz. **Bottom**: residual (in pixels) of the displacements from both methods.

pdf/101/2/415/3089332/joces_101_2_415.pdf.
[4] T. Schmidt, G. J. Schuetz, W. Baumgartner, H. J. Gruber, and H. Schindler, The Journal of Physical Chemistry **99**, 17662 (1995), https://doi.org/10.1021/j100049a030.
[5] R. Ghosh and W. Webb, Biophysical Journal **66**, 1301 (1994).
[6] S. M. Anthony, L. Hong, M. Kim, and S. Granick, Langmuir **22**, 9812 (2006).
[7] Y. Zhong, C. Li, H. Zhou, and G. Wang, Analytical Chemistry **90**, 10748 (2018).
[8] J. C. Crocker and D. G. Grier, Journal of Colloid and Interface Science **179**, 298 (1996).
[9] Y. Tang, F. Duan, A. Zhou, P. Kanitthamniyom, S. Luo, X. Hu, X. Jiang, S. Vasoo, X. Zhang, and Y. Zhang, Bioengineering & Translational Medicine **8**, e10428 (2022), https://aiche.onlinelibrary.wiley.com/doi/pdf/10.1002/btm2.10428.
[10] A. R. Wade and F. W. Fitzke, Optics Express **3**, 190 (1998).
[11] Y. Minowa, K. Kato, S. Ueno, T. W. Penny, A. Pontin, M. Ashida, and P. F. Barker, Review of Scientific Instruments **93** (2022).
[12] C. W. Lewandowski, T. D. Knowles, Z. B. Etienne, and B. D'Urso, Phys. Rev. Applied **15**, 014050 (2021), arXiv:2002.07585 [physics.app-ph].
[13] A. J. Berglund, M. D. McMahon, J. J. McClelland, and J. A. Liddle, Optics express **16**, 14064 (2008).
[14] M. K. Cheezum, W. F. Walker, and W. H. Guilford, Biophysical Journal **81**, 2378 (2001).
[15] Y. Feng, J. Goree, and B. Liu, Review of Scientific Instruments **78**, 053704 (2007).
[16] F. Melandsø, r. Bjerkmo, G. Morfill, H. Thomas, and M. Zuzic, Physics of Plasmas **7**, 4368 (2000).
[17] B. Liu, J. Goree, V. Nosenko, and L. Boufendi, Physics of Plasmas **10**, 9 (2003).
[18] Y. Ivanov and A. Melzer, Review of Scientific Instruments **78**, 033506 (2007).
[19] S.-H. Lee, Y. Roichman, G.-R. Yi, S.-H. Kim, S.-M. Yang, A. van Blaaderen, P. van Oostrum, and D. G. Grier, Opt. Express **15**, 18275 (2007).
[20] V. Svak, J. Flajšmanová, L. Chvátal, M. Šiler, A. Jonáš, J. Ježek, S. H. Simpson, P. Zemánek, and O. Brzobohatý, Optica **8**, 220 (2021).
[21] I. T. Leite, S. Turtaev, X. Jiang, M. Šiler, A. Cuschieri, P. S. J. Russell, and T. Čižmár, Nature Photonics **12**, 33 (2018).
[22] S. van der Walt, J. L. Schönberger, J. Nunez-Iglesias, F. Boulogne, J. D. Warner, N. Yager, E. Gouillart, T. Yu, and the scikit-image contributors, PeerJ **2**, e453 (2014).
[23] A. Paintdakhi, B. Parry, M. Campos, I. Irnov, J. Elf, I. Surovtsev, and C. Jacobs-Wagner, Molecular microbiology **99**, 767 (2016).
[24] J. M. Graving, D. Chae, H. Naik, L. Li, B. Koger, B. R. Costelloe, and I. D. Couzin, Elife **8**, e47994 (2019).
[25] B. H. Savitzky, S. E. Zeltmann, L. A. Hughes, H. G. Brown, S. Zhao, P. M. Pelz, T. C. Pekin, E. S. Barnard, J. Donohue, L. R. DaCosta, *et al.*, Microscopy and Microanalysis **27**, 712 (2021).
[26] A. Ö. Argunşah, E. Erdil, M. U. Ghani, Y. Ramiro-Cortés, A. F. Hobbiss, T. Karayannis, M. Çetin, I. Israely, and D. Ünay, Scientific Reports **12**, 12405 (2022).
[27] M. Guizar-Sicairos, S. T. Thurman, and J. R. Fienup, Optics letters **33**, 156 (2008).
[28] M. T. Hsu, V. Delaubert, P. K. Lam, and W. P. Bowen, Journal of Optics B: Quantum and Semiclassical Optics **6**, 495 (2004).
[29] B. R. Slezak, C. W. Lewandowski, J.-F. Hsu, and B. D'Urso, New Journal of Physics **20**, 063028 (2018).
[30] A. Pralle, M. Prummer, E.-L. Florin, E. Stelzer, and J. Hörber, Microscopy Research and Technique **44**, 378 (1999).
[31] J. Gieseler, B. Deutsch, R. Quidant, and L. Novotny, Phys. Rev. Lett. **109**, 103603 (2012).

[32] T. Li, S. Kheifets, and M. G. Raizen, Nature Physics **7**, 527 (2011).
[33] U. c. v. Delić, M. Reisenbauer, D. Grass, N. Kiesel, V. Vuletić, and M. Aspelmeyer, Phys. Rev. Lett. **122**, 123602 (2019).
[34] D. Windey, C. Gonzalez-Ballestero, P. Maurer, L. Novotny, O. Romero-Isart, and R. Reimann, Phys. Rev. Lett. **122**, 123601 (2019).
[35] Y. Arita, M. Mazilu, and K. Dholakia, Nature Communications **4**, 2374 (2013).
[36] Y. Arita, E. M. Wright, and K. Dholakia, Optica **5**, 910 (2018).
[37] Y. Arita, S. H. Simpson, P. Zemánek, and K. Dholakia, Science Advances **6**, eaaz9858 (2020), https://www.science.org/doi/pdf/10.1126/sciadv.aaz9858.
[38] Y. Ren, E. Benedetto, H. Borrill, Y. Savchuk, M. Message, K. O'Flynn, M. Rashid, and J. Millen, Applied Physics Letters **121**, 113506 (2022), https://pubs.aip.org/aip/apl/article-pdf/doi/10.1063/5.0106111/16484415/113506_1_online.pdf.
[39] G. Gallego, T. Delbrück, G. Orchard, C. Bartolozzi, B. Taba, A. Censi, S. Leutenegger, A. J. Davison, J. Conradt, K. Daniilidis, and D. Scaramuzza, IEEE Transactions on Pattern Analysis and Machine Intelligence **44**, 154 (2022).
[40] J.-F. Hsu, P. Ji, C. W. Lewandowski, and B. D'Urso, Scientific Reports **6**, 30125 (2016).
[41] W. M. Klahold, C. W. Lewandowski, P. Nachman, B. R. Slezak, and B. D'Urso, in *Optical, Opto-Atomic, and Entanglement-Enhanced Precision Metrology*, Vol. 10934, edited by S. M. Shahriar and J. Scheuer, International Society for Optics and Photonics (SPIE, 2019) p. 109340P.
[42] L. R. Werneck, https://github.com/leowerneck/RETINAS (Visited on: October 12, 2023).
[43] L. R. Werneck, C. Jessup, A. Brandenberger, T. Knowles, C. W. Lewandowski, M. Nolan, K. Sible, Z. B. Etienne, and B. D'Urso, Zenodo, doi:10.5281/zenodo.8336549 (2023).